\documentclass[doublecol]{epl2} 
\usepackage{nicefrac}

\usepackage{times}
\usepackage{xcolor}
\usepackage{amsmath}
\usepackage{epsfig}
\usepackage{dcolumn}
\usepackage{amssymb}

\title{Spin freezing and the Sachdev-Ye model}

\author{Philipp Werner,\inst{1} Aaram J. Kim,\inst{2} and Shintaro Hoshino\inst{3}}
\shortauthor{Philipp Werner, Aaram Kim and Shintaro Hoshino}

\institute{                    
  \inst{1} Department of Physics, University of Fribourg, 1700 Fribourg, Switzerland\\
  \inst{2} Department of Physics, King's College London, Strand, London WC2R 2LS, UK\\
  \inst{3} Department of Physics, Saitama University, Saitama 338-8570, Japan
}
\pacs{71.10.Fd}{}

\abstract{
Spin-freezing is the origin of bad-metal physics and non-Fermi liquid (non-FL) properties in a broad range of correlated compounds. 
In a multi-orbital lattice system with Hund coupling, doping of the half-filled Mott insulator results in a highly incoherent metal with frozen magnetic moments. 
These moments fluctuate and collapse in a crossover region that is characterized by unusual non-Fermi liquid properties such as a self-energy whose imaginary part varies  
$\propto \sqrt{\omega}$ over a significant energy range. At low enough temperature,  
the local moment fluctuations induce electron pairing,  
{which may be a generic mechanism for}
unconventional superconductivity. 
While this physics has been discovered in numerical studies of multi-orbital Hubbard systems, it exhibits a striking similarity to the analytically solvable Sachdev-Ye (SY) model, 
and its recent fermionic extensions. 
Here, we 
{explore} 
the relation between spin-freezing and SY physics, and thus shed light on fundamental properties of Hund metals. 
}

\begin{document}

\maketitle

\section{Introduction}
\begin{figure*}[t]
\begin{center}
\includegraphics[angle=0, width=0.75\textwidth]{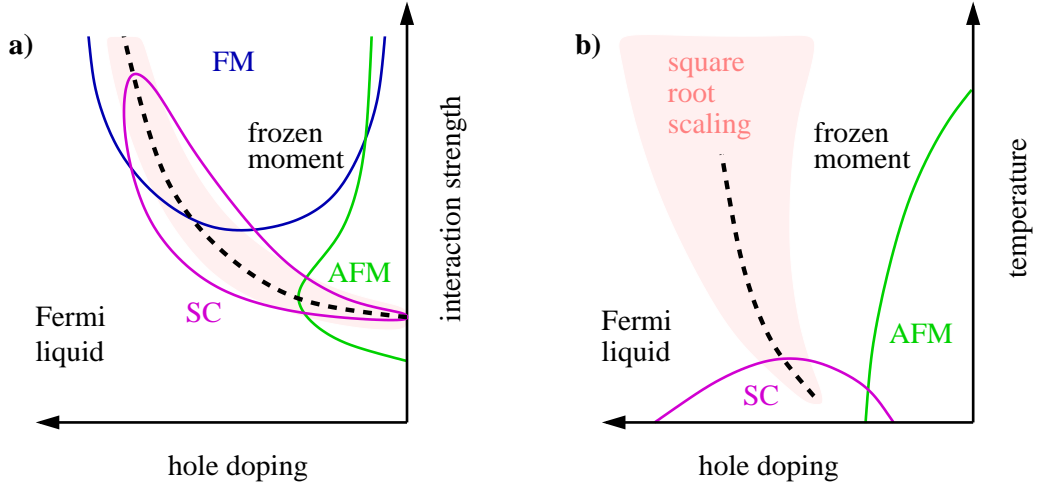}
\caption{Sketch of the generic spin-freezing behavior observed in multi-orbital Hubbard systems with Hund coupling. Panel a) plots the phase diagram at 
{a low, but nonzero}
temperature, while panel b) shows the result for fixed interaction strength. The black dashed line indicates the spin-freezing crossover line in the symmetric phase, while the light-pink shading covers the region in which a characteristic non-FL behavior with $\text{Im}\Sigma(\omega)\propto \sqrt{\omega}$ is observed. 
Potential instabilities to antiferromagnetic (AFM), ferromagnetic (FM) and orbital-singlet spin-triplet superconductivity (SC) are indicated by green, blue and violet lines. As temperature is lowered further, the SC region expands, while the FM and AFM regions remain almost unchanged.  
}
\label{fig_illustration}
\end{center}
\end{figure*}

{The discovery of spin-freezing and associated non-FL properties in dynamical mean field theory (DMFT) \cite{Georges1996} investigations of multi-orbital Hubbard systems \cite{Werner2008,Haule2009,Ishida2010} led to the concept of ``Hund metals," which plays an important role in theoretical studies of unconventional superconductors and other correlated compounds \cite{Georges2013}. 
These phenomena are 
associated with locally fluctuating composite spins, and are}
characteristic of strongly interacting multi-orbital systems with Hund coupling, 
since they appear 
in two- \cite{Medici2011,Hafermann2012,Hoshino2016,Steiner2016,Werner2016}, three- \cite{Werner2008,Werner2009,Medici2011,Yin2012,Hoshino2015,Hung2015,Stadler2015,Hoshino2016}, and five-orbital systems \cite{Haule2009,Ishida2010,Liebsch2010,Werner2012,Pelliciari2017}, in models with density-density \cite{Ishida2010,Hafermann2012,Hoshino2015,Hoshino2016} or spin-rotation invariant \cite{Werner2008,Medici2011,Hoshino2015,Stadler2015,Hoshino2016} interactions, and in spin-orbit coupled systems \cite{Kim2017,Kim2018}.    
Recent studies \cite{Hoshino2015,Hoshino2016,Steiner2016} furthermore revealed a superconducting instability in the spin-freezing crossover regime. The resulting 
phase diagrams exhibit the generic features expected of unconventional superconductors, namely a superconducting dome next to a magnetically ordered phase and a non-FL metal above the superconducting dome \cite{Hoshino2015,Steiner2016}, which crosses over to a more conventional metal as doping is increased or the interaction is reduced (Fig.~\ref{fig_illustration}).
These results are relevant for the understanding of unconventional superconductivity in materials with heavy elements, such as strontium ruthenate compounds and uranium-based superconductors \cite{Hoshino2015}. Alkali-doped fulleride compounds \cite{Capone2009,Nomura2012} exhibit the same phenomenology \cite{Steiner2016,Hoshino2017}, but with the roles of spin and orbital degrees of freedom interchanged because of the effectively negative Hund coupling. Also the cuprate phase diagram has a natural interpretation within the spin-freezing picture \cite{Werner2016},
{while in the case of iron pnictides, at least the normal state properties are strongly influenced by a spin-freezing crossover \cite{Haule2009, Ishida2010, Werner2012}.}

This physics has 
not yet received the proper attention outside of the DMFT community. One reason may be that the spin-freezing and non-FL behavior emerges from a numerical multi-orbital impurity calculation, which at first sight seems inaccessible to simple semi-analytical treatments. At the same time, the DMFT results exhibit a 
remarkable similarity to the physics of the analytically solvable Sachdev-Ye (SY) model \cite{Sachdev1993}, i.e. the large-$N$ limit of an infinitely connected random Heisenberg model of SU($N$) spins. This model yields a non-FL self-energy with Im$\Sigma\propto\sqrt{\omega}$, which arises from the disordering of a spin-glass state by quantum fluctuations. Parcollet and Georges \cite{Parcollet1999} studied a doped version of the SY model (a disordered SU$(N)$ $t$-$J$ model) and obtained a temperature-doping phase diagram with close resemblance to the generic spin-freezing phase diagram sketched in Fig.~\ref{fig_illustration}. 
In the meantime a fermionic version of the SY model, dubbed Sachdev-Ye-Kitaev (SYK) model has been introduced \cite{Kitaev2015,Sachdev2015}, and very recently, Chowdhury and co-workers \cite{Chowdhury2018} formulated and analyzed a model of translationally invariant ``SYK dots." The properties of this model are again qualitatively similar to those observed in multi-orbital DMFT simulations, and the set-up of Ref.~\cite{Chowdhury2018} bears more resemblance to the 
systems considered in the DMFT studies, altough the physical significance of a Gaussian-distributed interaction tensor remains unclear. 

\begin{figure*}[t]
\begin{center}
\includegraphics[angle=-90, width=0.39\textwidth]{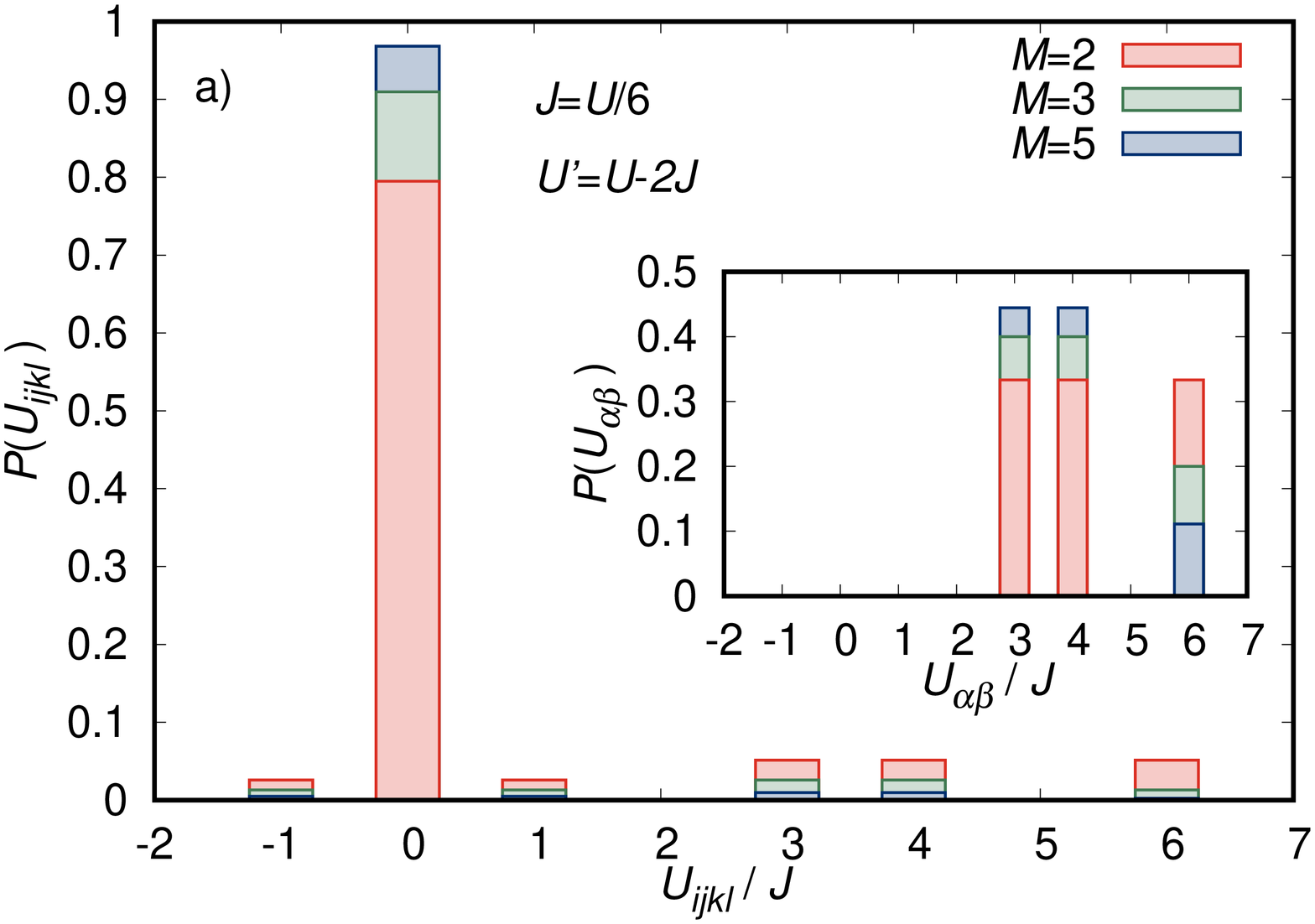} \hspace{10mm}
\includegraphics[angle=-90, width=0.39\textwidth]{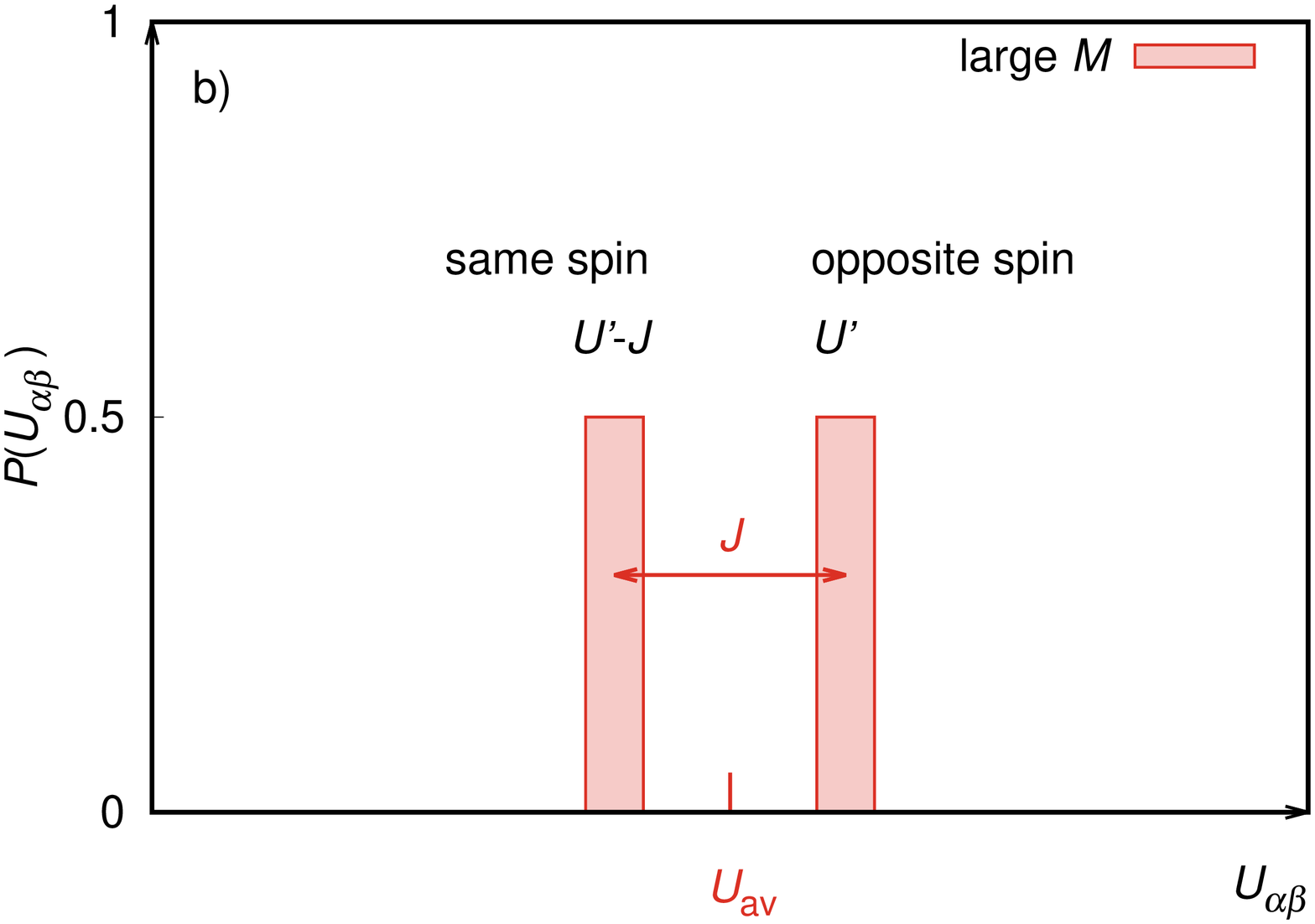}
\caption{
Distribution of values of the interaction tensor for an $M$-orbital Hubbard model. 
Panel a): Slater-Kanamori interaction. The main panel shows the result for the spin-rotation invariant model parametrized by $H_{\rm int}=\sum_{ijkl} U_{ijkl}c^\dagger_ic^\dagger_jc_kc_l$ and the inset the result for the density-density approximation $H_\text{int}=\sum_{\alpha\beta}U_{\alpha\beta}n_\alpha n_\beta$.
Panel b): Bimodal probability distribution $P(U_{\alpha\beta})$ representing inter-orbital interactions.  
}
\label{fig_dist}
\end{center}
\end{figure*}

\section{Results}
To {shed some light on} the relation between spin-freezing and SYK physics we introduce a simple lattice model which captures the essential features of multi-orbital Hubbard systems, and allows us to interpret the results of Refs.~\cite{Sachdev1993,Parcollet1999,Chowdhury2018} in terms of Hund coupling effects.      
We start with a brief analysis of the distribution of interaction values in an $M$-orbital Hubbard model with Slater-Kanamori interaction $H_\text{int}=\sum_{ijkl} U_{ijkl}c^\dagger_ic^\dagger_jc_kc_l$ parametrized by the intra-orbital interaction $U$, inter-orbital interaction $U'$ and the Hund coupling $J$. Here, we use flavor indices which combine orbital and spin degrees of freedom ($N=2M$ flavors). The possible values of the interaction tensor $U_{ijkl}$ and their distribution are listed in Tab.~\ref{tab} and illustrated in Fig.~\ref{fig_dist}a). 
In the large-$M$ limit the relevant nonzero terms are the density-density interactions between electrons in different orbitals, with energies $U'$ and $U'-J$ (depending on the relative orientation of the spin), as well as the spin-flip and pair-hopping terms. Since spin-freezing physics is already observed in models with density-density interactions \cite{Medici2011,Hafermann2012,Hoshino2015}, we focus on the density-density approximation 
$H_\text{int}=\sum_{\alpha<\beta}U_{\alpha\beta}n_\alpha n_\beta$, for which the distribution of interaction values is plotted in the inset of Fig.~\ref{fig_dist}a). For a large number of orbitals, it is sufficient to keep the two values of $U_{\alpha\beta}$ which represent interorbital interactions, and which can be parametrized by the average interaction $U_\text{av}$ and the Hund coupling $J$ (Fig.~\ref{fig_dist}b)).

\begin{table*}
\begin{center}
\begin{tabular}{lll}
$U_{ijkl}$ & value & number of independent terms \\
\hline
intra-orbital & $U$	& $M$ \\
inter-orbital (opposite spin) \hspace{5mm}& $U'$ & $M(M-1)$ \\
inter-orbital (same spin) & $U'-J$ \hspace{5mm} & $M(M-1)$ \\
spin-flip	& $-J$ & $M(M-1)/2$ \\
pair-hopping	& $J$ & $M(M-1)/2$ \\
other		& 0 & $2M^4 - 2M^3 - \tfrac{3}{2}M^2 + \tfrac{3}{2}M$
\end{tabular}
\end{center}
\caption{Values of the interaction tensor in an $M$-orbital system with Slater-Kanamori interaction.}
\label{tab}
\end{table*}

In the spirit of the SY model, we choose these interactions randomly among the two values $U_{\alpha\beta} = \frac{\tilde U}{N} \pm \frac{\tilde J}{\sqrt{N}}$, 
i.e. using the probability distribution 
\begin{align}
P(U_{\alpha\beta}) =& \tfrac{1}{2}\delta\big(U_{\alpha\beta}-[\tilde U/N + \tilde J/\sqrt{N}]\big) \nonumber\\
 &\quad + \tfrac{1}{2}\delta\big(U_{\alpha\beta}-[\tilde U/N - \tilde J/\sqrt{N}]\big).
\label{p_bimodal}
\end{align} 
{This random inter-orbital interaction mimics the effect of local spin fluctuations in the spin-freezing crossover regime.}
$\tilde U$ represents the monopole interaction and $\tilde J$ the strength of the Hund coupling.  The constraint for physical (repulsive) interactions is $N<(\tfrac{\tilde U}{\tilde J})^2$.

Our {translation invariant} lattice system contains $N$ flavors per site, local density-density interactions and flavor-diagonal hoppings 
{$t_{rr'}$}
between different sites,
\begin{equation}
H= \sum_{rr'} \sum_{\alpha} (t_{rr'} - \mu \delta_{rr'})c^\dagger_{r\alpha}c_{r'\alpha} + \sum_r\sum_{\alpha<\beta} U_{\alpha\beta} n_{r\alpha}n_{r\beta},\label{eq_lattice}
\end{equation}
where the interactions $U_{\alpha\beta}$ are site independent and distributed according to Eq.~(\ref{p_bimodal}).  
{The Fourier transform of $t_{rr'}$ gives the dispersion $\epsilon_k$.}
We can directly average the partition function over the 
bimodal distribution of $U_{\alpha\beta}$ to obtain $Z \propto \text{Tr}_c \big[ e^{-S_0-S_\text{int}}\big],$ with
\begin{align}
S_0 = & \int d\tau \sum_r \sum_\alpha c^\dagger_{r\alpha} (\partial_\tau - \mu)c_{r\alpha}\nonumber\\
 &+ \int d\tau \sum_{rr'} \sum_\alpha t_{rr'} (c^\dagger_{r\alpha}c_{r'\alpha}+c^\dagger_{r'\alpha}c_{r\alpha}),\label{eq_S0}\\
S_\text{int} = & \frac{\tilde U}{2N} \int d\tau \sum_{r} \Big(\sum_\alpha n_{r\alpha}\Big)^2 \nonumber\\
 & - \frac{\tilde J^2}{4N} \int d\tau d\tau' \sum_{rr'} \Big( \sum_\alpha n_{r\alpha}(\tau) n_{r'\alpha}(\tau') \Big)^2.\label{eq_Sint} 
\end{align}
To integrate out the fermions 
we introduce $G_{r'r\alpha}(\tau',\tau)=c^\dagger_{r\alpha}(\tau)c_{r'\alpha}(\tau')$ 
and 
write the interaction term as 
$S_\text{int} = \frac{\tilde U}{2N} \int d\tau \sum_{r} \big(\sum_\alpha G_{rr\alpha}(\tau,\tau)\big)^2 
- \frac{\tilde J^2}{4N} \int d\tau d\tau' \sum_{rr'} \big(\sum_\alpha G_{rr'\alpha}(\tau,\tau') G_{r'r\alpha}(\tau',\tau) \big)^2$, 
supplemented by the contraints (for all $r,r',\alpha$) 
$1=\int DG_{rr'\alpha}$ $\times \int D\Sigma_{rr'\alpha} e^{\int d\tau d\tau' \Sigma_{rr'\alpha}(\tau,\tau') (G_{r'r\alpha}(\tau',\tau)-c^\dagger_{r\alpha}(\tau)c_{r'\alpha}(\tau'))}$.
Integrating out the fermions in the presence of the constraining fields results in the partition function $Z \propto \big\{\prod D G_{rr'\alpha} D \Sigma_{rr'\alpha}\big\} e^{-S_\text{eff}[\Sigma,G]}$ with
\begin{align}
& S_\text{eff}[\Sigma,G] = -\sum_\alpha \text{Tr} \ln [\partial_\tau - \mu + \epsilon_k+\Sigma_\alpha] \nonumber\\
&\quad + \frac{\tilde U}{2N} \int d\tau \sum_{r} \Big(\sum_\alpha G_{rr\alpha}(\tau,\tau)\Big)^2 \nonumber\\
&\quad - \frac{\tilde J^2}{4N} \int d\tau d\tau' \sum_{rr'} \Big(\sum_\alpha G_{rr'\alpha}(\tau,\tau') G_{r'r\alpha}(\tau',\tau) \Big)^2\nonumber\\
&\quad -\int d\tau d\tau' \sum_{rr'}\sum_\alpha \Sigma_{rr'\alpha}(\tau,\tau') G_{r'r\alpha}(\tau',\tau) .
\end{align}
At the saddle point of this action, we have $G_{rr'\alpha}\equiv G_{rr'}$ and $\Sigma_{rr'\alpha}\equiv \Sigma_{rr'}$, because of the orbital degeneracy 
and the saddle point equations $\delta Z/\delta G_{r'r}(\tau',\tau)=0$ and $\delta Z/\delta \Sigma_{r'r}(\tau',\tau)=0$ yield
\begin{align}
&G_{k}^{-1}(i\omega_n) = i\omega_n+\mu-\epsilon_k-\Sigma_k(i\omega_n),\label{inv_Gk}\\
&\Sigma_{rr'}(\tau) =\tilde U \delta_{rr'}\delta(\tau)G_{rr'}(\tau^-)
\!-\! \tilde J^2 G_{rr'}(\tau) G_{r'r}(\!-\tau)  G_{rr'}(\tau).\label{IPT}
\end{align}
In 
deriving these equations, 
we neglected fluctuations in the transverse direction, i.e. orthogonal to the subspace defined by $G_{rr'\alpha}\equiv G_{rr'}$ and $\Sigma_{rr'\alpha}\equiv \Sigma_{rr'}$.  
{Also, the saddle point solution is an approximate solution even in the large-$N$ limit, because the identification of $G_{r'r\alpha}(\tau',\tau)$ with $c^\dagger_{r\alpha}(\tau)c_{r'\alpha}(\tau')$ is only valid at the level of expectation values, and not individual paths.}

\begin{figure*}[t]
\begin{center}
\includegraphics[angle=0, width=0.85\textwidth]{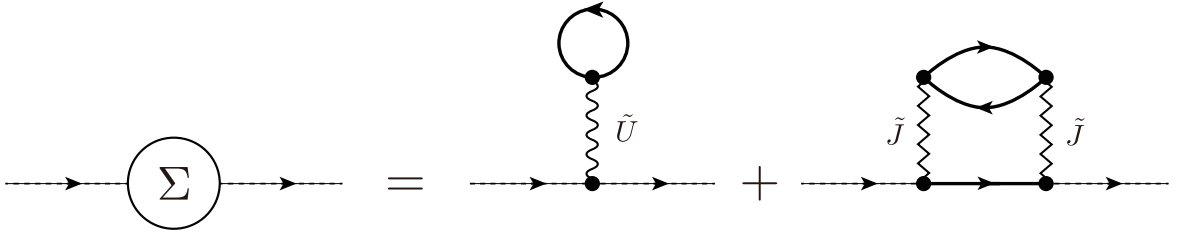}
\caption{Schematic self-energy in the large-$N$ paramagnetic limit. Solid lines represent the interacting Green function.
}
\label{fig_diag}
\end{center}
\end{figure*}

The self-energy (\ref{IPT}) is sketched in Fig.~\ref{fig_diag}. While this self-energy with Hartree and second-order diagram at first sight looks similar to the result for a single-orbital Hubbard model in self-consistent second-order perturbation theory, there are important differences. First of all, 
the strength of the second order term is controlled by the Hund coupling $\tilde J$, while the monopole interaction $\tilde U$ determines the Hartree shift. Without Hund coupling, there is no interesting non-FL behavior, which is consistent with the results from DMFT studies. Second, this self-energy is not the result of a truncation of some weak-coupling expansion, but the result obtained for a strongly correlated lattice model in the limit of a large number of orbitals.

Since the self-energy (\ref{IPT}) is (up to the Hartree term) 
identical to the one discussed in Ref.~\cite{Chowdhury2018}, the analysis of the non-FL properties of this theory is completely analogous. The system is a Fermi liquid for $T \ll T^*\approx W^2/\tilde J$, where $W$ is the bandwidth of the noninteracting model, while for large enough $\tilde J$, there is a temperature range $T^*\ll T \ll \tilde J$ in which the system exhibits a non-FL self-energy analogous to the (single-site) SY model \cite{Sachdev1993,Parcollet1999}. For $T \gg \tilde J$ the Hund coupling is no longer active and the local moment fluctuates freely. 

To understand the non-FL behavior, we absorb the Hartree term in Eq.~(\ref{IPT}) into the chemical potential by the shift $\mu\rightarrow \tilde \mu=\mu-\tilde Un$. We further assume that the energy is high enough ($\omega_n\gg W^2/\tilde J$) that we can neglect the dispersion, but small enough ($\omega_n\ll\tilde{J}$) that the $i\omega_n$ term remains irrelevant. In this case the problem becomes local ($\Sigma_k=\Sigma, G_{rr'}=\delta_{rr'}G$) and formally identical to the problem studied by Sachdev and Ye \cite{Sachdev1993},  
\begin{align}
G^{-1}(i\omega_n) &= i\omega_n+\tilde\mu-\Sigma(i\omega_n),\\
\Sigma(\tau) &= -\tilde J^2 G(\tau) G(-\tau) G(\tau). 
\end{align}
The particle-hole symmetric solution ($\tilde \mu=0$) 
for \mbox{$T^* \ll \omega_n \ll \tilde J$}
is\cite{Sachdev2015} 
\begin{align}
	G(i\omega_n) 
=& -i\Big(\frac{2\pi}{\tilde J^2}\Big)^{1/4}\frac{1}{\sqrt{\omega_n}}+\ldots, \\
\Sigma(i\omega_n)=&
-i\Big(\frac{\tilde J^2}{2\pi}\Big)^{1/4}\sqrt{\omega_n} + \dots , 
\label{eq_sigma}
\end{align}
which is consistent with the leading-order expression 
for $G^{-1}(i\omega_n)=-\Sigma(i\omega_n)$.

\section{Discussion}
While it is remarkable that the large-$N$ analysis of our simple model predicts the characteristic non-FL behavior of realistic two-, three- and five-orbital Hubbard systems, we also have to point out some differences. For example, the FL coherence scale in multi-orbital Hubbard systems with Hund coupling is very low \cite{Stadler2015},  and not compatible with the simple estimate $T^*=W^2/J$ with bare bandwidth $W$ and Hund coupling $J$. 
The Hund coupling leads to the formation of composite moments with large spin, and the FL state emerges when these moments are screened below a $T^*_\text{Kondo}$ that is
exponentially suppressed with the size of the spin \cite{Nevidomskii2009,Georges2013}.  
Numerical simulations suggest $T^* \approx (ZW)^2/J$, where $Z$ is the quasi-particle weight of the low-temperature Fermi liquid and $J$ the bare Hund coupling \cite{Stadler2015}.  
Another difference is that in DMFT, the non-FL behavior only appears for large $U$, upon doping of the half-filled Mott insulator. The $\text{Im}\Sigma(i\omega_n)\sim\sqrt{\omega_n}$ scaling is found in a crossover region between an incoherent metal state with frozen magnetic moments ($\text{Im}\Sigma(i\omega_n\rightarrow 0)\sim \text{const}>0$) \cite{Werner2008}, and a FL metal phase, see Fig.~\ref{fig_illustration}. This indicates that quantum fluctuations in the large-$N$ description are overestimated. (It is interesting to note that the bosonic representation of the SY model yields a ``local moment" phase with spin correlation functions reminiscent of the spin-frozen regime for large enough magnitude of the spin \cite{Georges2001}.)  

Without averaging over the interactions $U_{\alpha\beta}$, the lattice system (\ref{eq_lattice}) becomes local in the strong coupling limit and does not exhibit any nontrivial electronic fluid behavior. The non-locality with respect to imaginary time in Eq.~(\ref{eq_Sint}), and hence the second order diagram in the self-energy (\ref{IPT}) are generated by the interaction part in the averaging process. This mimics the retardation originating from the intersite  
hopping in the original multi-orbital Hubbard model, and effectively locates the system 
in the spin-freezing crossover regime. 
$\tilde J$ should thus be regarded as a renormalized parameter which includes inter-site hopping effects, and also the dispersion $\epsilon_k$ in Eq.~(\ref{inv_Gk}) and the corresponding bandwidth $W$ (which determines the crossover scale $T^*=W^2/\tilde J$) are renormalized parameters.

Based on the above discussion, we propose the following interpretation of the generic spin-freezing phenomenology (Fig.~\ref{fig_illustration}): As the filling or interaction in the multi-orbital system is increased, local moments appear in the metal phase due to the Hund coupling. As these moments form, the FL coherence temperature $T^*$ drops and the system enters into a spin-frozen metal state (away from half filling) or into a Mott phase (at half filling). In the crossover regime to the spin-frozen state the moments are slowly fluctuating, so that the SY-type non-FL behavior emerges for $T^*\lesssim T \lesssim \tilde J$. The large-$N$ analysis describes the filling and interaction range in which local moments are present, but fluctuations prevent the freezing of these moments. 

Finally we would like to comment on the electronic ordering tendencies. Recent DMFT simulations\cite{Hoshino2015,Hoshino2016} revealed that the spin-frozen regime near half-filling is prone to antiferromagnetic order, while at large interactions and large doping, the system tends to order ferromagnetically. Most interestingly, along the spin-freezing crossover line, there is an instability to (spin-triplet) superconductivity.  
To explain the latter we 
define the effective interaction $U_\text{eff}$ which takes into account the effect of the ``polarization bubble" $P(\tau)=G(\tau)G(-\tau)$,
\begin{equation}
U_\text{eff}(i\omega_n) = \tilde U + \tilde J P(i\omega_n) \tilde J.
\end{equation}
If $\text{Re}U_\text{eff}(i\omega_n\rightarrow 0)$ becomes negative enough, it should induce a pairing between electrons in different orbitals. From Eq.~(\ref{eq_sigma}) we find $P(\tau)=-\frac{1}{\tilde J\tau}\frac{1}{\sqrt{2\pi}} $ and therefore  
\begin{equation}
P(i\omega_n)=-\frac{1}{\sqrt{2\pi}}\frac{1}{\tilde J}[\log(\tilde J/\omega_n)-\gamma], \quad W^2/\tilde J \ll \omega_n \ll \tilde J. 
\end{equation}
Hence, $\text{Re}P(i\omega_n\rightarrow 0) \rightarrow -\infty$ and the pairing indeed occurs if  
the attractive interaction is realized above $T^*$.  
Since the interaction favors ``high-spin" states (Fig.~\ref{fig_dist}b)), the pairing is ``spin triplet." 
{We note that pairing induced by enhanced local spin or orbital fluctuations has been discussed in connection with different classes of unconventional superconductors \cite{Hoshino2015,Steiner2016,Hoshino2017}, including cuprates \cite{Werner2016}, and may be a unifying principle for unconventional superconductivity. }
 
{The arguments presented in this paper are not a rigorous proof, but they suggest that the SY equations can be regarded as an effective description of} 
Hund-coupling induced non-FL properties of doped multi-orbital Hubbard systems, and they underscore the deep connection between spin-freezing and unconventional  superconductivity. Remarkably, the non-FL properties of the large-$N$ limit leave clear traces already in two- and three-orbital Hubbard systems, if the filling and interaction is tuned to a region where local moment fluctuations prevail.

\acknowledgments

PW acknowledges support from ERC Consolidator Grant 724103. AJK was supported by EPSRC through grant EP/P003052/1.

\bibliographystyle{eplbib}

\end{document}